\documentclass[a4paper]{jpconf}
\usepackage{graphicx}
\begin{document}
\title{Phase transitions in exactly solvable decorated model 
of localized Ising spins and itinerant electrons}

\author{J Stre\v{c}ka$^1$, A Tanaka$^2$, M Ja\v{s}\v{c}ur$^1$}

\address{$^1$ Department of Theoretical Physics and Astrophysics, Faculty of Science, \\ 
P. J. \v{S}af\'{a}rik University, Park Angelinum 9, 040 01 Ko\v{s}ice, Slovak Republic}
\address{$^2$ Department of General Education, Ariake National College of Technology, \\
Omuta, Fukuoka 836-8585, Japan}

\ead{jozef.strecka@upjs.sk}

\begin{abstract}
A hybrid lattice-statistical model of doubly decorated two-dimensional lattices, which have 
localized Ising spins at its nodal sites and itinerant electrons delocalized over decorating 
sites, is exactly solved with the help of a generalized decoration-iteration transformation. Under the assumption of a quarter filling of each couple of the decorating sites, 
the ground state constitutes either spontaneously long-range ordered ferromagnetic or ferrimagnetic 
phase in dependence on whether the ferromagnetic or antiferromagnetic interaction between 
the localized Ising spins and itinerant electrons is considered. The critical temperature of 
the spontaneously long-range ordered phases monotonically increases upon strengthening
the ratio between the kinetic term and the Ising-type exchange interaction.  
\end{abstract}

\section{Introduction}
Exactly soluble lattice-statistical models traditionally attract appreciable scientific interest 
as they offer a valuable insight into diverse aspects of cooperative phenomena \cite{wu09}. It should 
be mentioned, however, that sophisticated mathematical methods must be usually employed when 
searching for an exact treatment of even relatively simple interacting many-body systems and consequently, a theoretical treatment of more realistic or more complex models is often 
accompanied with a substantial increase of computational difficulties. The mapping technique 
based on generalised algebraic transformations belongs to the simplest mathematical 
methods, which allow to obtain the exact solution of a more complicated model from a precise 
mapping relationship with a simpler exactly solved model \cite{fish59,roja09}. Recently, 
this approach has been applied to an intriguing diamond chain model of interacting spin-electron 
system \cite{pere08,pere09}. The purpose of this work is to treat exactly similar two-dimensional 
(2D) model, which should provide a deeper insight into phase transitions and critical phenomena 
of interacting spin-electron systems.   

\section{Model and its exact solution}
\begin{figure}[h]
\includegraphics[width=20pc]{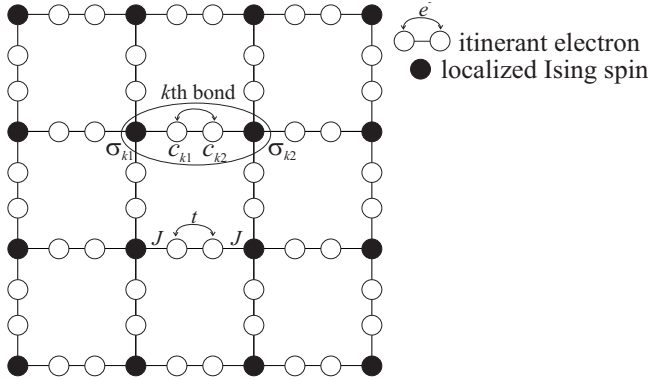}\hspace{0pc}%
\begin{minipage}[b]{18pc}\caption{\label{label}A part of the doubly decorated square lattice, 
which has at each nodal site (full circle) one localized Ising spin and at each couple of decorating sites (empty circles) one itinerant electron. The ellipse demarcates unit cell described 
by the $k$th bond Hamiltonian given by Eq.~(\ref{ham}).}
\label{fig1}
\end{minipage}
\end{figure}

Let us consider a hybrid lattice-statistical model of interacting spin-electron system 
on doubly decorated 2D lattices, which have one localized Ising spin at each nodal site 
and one itinerant electron delocalized over each couple of decorating sites. The magnetic 
structure of the model under investigation is schematically illustrated in figure \ref{fig1} 
on the particular example of the doubly decorated square lattice.  
For further convenience, let us define the total Hamiltonian as a sum over bond Hamiltonians 
$\hat{\cal H} = \sum_k \hat{\cal H}_k$, where each bond Hamiltonian $\hat{\cal H}_k$ involves 
all the interaction terms of one itinerant electron from the $k$th couple of the decorating sites
\begin{eqnarray}
\hat{\cal H}_k = \! \! \! &-& \! \! \! t \left( c^{\dagger}_{k1, \uparrow} c^{}_{k2, \uparrow} 
                          + c^{\dagger}_{k1, \downarrow} c^{}_{k2, \downarrow}
                          + c^{\dagger}_{k2, \uparrow} c^{}_{k1, \uparrow}  
                          + c^{\dagger}_{k2, \downarrow} c^{}_{k1, \downarrow} \right)	\nonumber \\
                 \! \! \! &-& \! \! \! 
          \frac{J}{2} \left[ \hat{\sigma}_{k1}^z \left( c^{\dagger}_{k1, \uparrow} c^{}_{k1, \uparrow} 
                                        -  c^{\dagger}_{k1, \downarrow} c^{}_{k1, \downarrow} \right) 
               + \hat{\sigma}_{k2}^z \left( c^{\dagger}_{k2, \uparrow} c^{}_{k2, \uparrow} 
                                -  c^{\dagger}_{k2, \downarrow} c^{}_{k2, \downarrow} \right) \right]. \label{ham}
\end{eqnarray}
In above, $c^{\dagger}_{k \alpha, \gamma}$ and $c^{}_{k \alpha, \gamma}$ ($\alpha = 1,2$, 
$\gamma = \uparrow, \downarrow$) denote usual creation and annihilation fermionic operators 
and $\hat{\sigma}_{k \alpha}^z$ is the standard spin-1/2 operator with the eigenvalues 
${\sigma}_{k \alpha}^z = \pm 1/2$. The transfer integral $t$ takes into account a kinetic energy 
of a single electron delocalized over a couple of the decorating sites and the exchange integral 
$J$ describes the Ising-type interaction between the delocalized electron and its two nearest-neighbouring localized Ising spins.

The crucial step of our calculation represents an evaluation of the partition function. 
A validity of the commutation relation between different bond Hamiltonians
$[\hat{\cal H}_i, \hat{\cal H}_j] = 0$ allows a straightforward factorization 
of the total partition function ${\cal Z}$ into a product performed 
over bond partition functions ${\cal Z}_k$
\begin{eqnarray}
{\cal Z} = \mbox{Tr}_{\{ \sigma_i \}} \mbox{Tr}_{\{ c_i \}} \exp \left(- \beta \hat{\cal H} \right)
           = \mbox{Tr}_{\{ \sigma_i \}} \prod_{k=1}^{Nq/2} 
             \mbox{Tr}_{c_{k1}, c_{k2}} \exp \left(- \beta \hat{\cal H}_k \right)
           = \mbox{Tr}_{\{ \sigma_i \}} \prod_{k=1}^{Nq/2} {\cal Z}_k, 
\label{pf}
\end{eqnarray}
where $\beta = 1/(k_{\rm B} T)$, $k_{\rm B}$ is Boltzmann's constant, $T$ is the absolute temperature,
$N$ is the total number of the Ising spins (i.e. the nodal lattice sites) and $q$ is their coordination number (i.e. the number of nearest neighbours). Next, the symbols $\mbox{Tr}_{\{ \sigma_i \}}$ and $\mbox{Tr}_{\{ c_i \}}$ denote a trace over degrees of freedom of all Ising spins and itinerant electrons, respectively, while the symbol $\mbox{Tr}_{c_{k1}, c_{k2}}$ stands for a trace over 
degrees of freedom of the itinerant electron from the $k$th couple of decorating sites. 
After the elementary diagonalisation of the bond Hamiltonian $\hat{\cal H}_k$, one arrives 
at the explicit expression of the bond partition function ${\cal Z}_k$, which can be eventually 
replaced through the appropriately chosen generalised decoration-iteration transformation \cite{fish59,roja09}
\begin{eqnarray}
{\cal Z}_k = 4 \cosh \left[ \frac{\beta J}{4} (\sigma_{k1}^z + \sigma_{k2}^z) \right] 
   \cosh \left[ \frac{\beta}{4} \sqrt{J^2 (\sigma_{k1}^z - \sigma_{k2}^z)^2 + (4t)^2} \right]
   = A \exp(\beta R \sigma_{k1}^z \sigma_{k2}^z).
\label{dit}
\end{eqnarray} 
The 'self-consistency' condition of the algebraic transformation (\ref{dit}) requires that this 
mapping relationship must hold independently of the spin states of two Ising spins $\sigma_{k1}^z$ 
and $\sigma_{k2}^z$ involved therein, which directly determines so far not specified transformation parameters $A$ and $R$ as
\begin{eqnarray}
A = 4 \left \{\cosh \left( \beta t \right) \cosh \left( \frac{\beta J}{4} \right) 
            \cosh \left[ \frac{\beta}{4} \sqrt{J^2 + (4t)^2} \right] \right \}^{1/2} \! \! \!,  \quad 
\beta R = 2 \ln \left \{ \frac{\cosh \left( \beta t \right) \cosh \left( \frac{\beta J}{4} \right) } 
                          {\cosh \left[ \frac{\beta}{4} \sqrt{J^2 + (4t)^2} \right]} \right \}\!.
\label{mp}
\end{eqnarray}  
At this stage, a substitution of the generalised decoration-iteration transformation (\ref{dit}) into Eq.~(\ref{pf}) yields in turn a simple mapping relation between the partition function of the interacting spin-electron system on the doubly decorated 2D lattice and the partition function 
of the simple spin-1/2 Ising model on the corresponding undecorated lattice
\begin{eqnarray}
{\cal Z} (\beta, J, t) = A^{Nq/2} {\cal Z}_{{\rm IM}} (\beta, R). 
\label{pfex}
\end{eqnarray} 
It can be easily understood from Eqs.~(\ref{mp}) and (\ref{pfex}) that the mapping parameter $A$ 
cannot cause a non-analytic behaviour of the partition function ${\cal Z}$ and hence, the investigated spin-electron system becomes critical if and only if the corresponding spin-1/2 Ising model becomes critical as well. Accordingly, the lines of critical points can readily be obtained from a comparison of the effective temperature-dependent coupling $\beta R$ with the relevant critical points of the spin-1/2 Ising model on the corresponding undecorated lattices that are given by $\beta_{\rm c} R = 2 \ln(2 + \sqrt{3})$, $2 \ln (1 + \sqrt{2})$, and $\ln 3$ for the honeycomb, square, 
and triangular lattices \cite{lin92}, respectively. 

Other thermodynamic quantities can be now easily derived from the mapping relation (\ref{pfex})
between the partition functions. For instance, both sublattice magnetisations can be calculated by combining the mapping relation (\ref{pfex}) with the exact Callen-Suzuki identity. As a result, 
the sublattice magnetisations $m_i$ and $m_e$ per one Ising spin and per one itinerant electron read 
\begin{eqnarray}
m_i = m_{{\rm IM}} (\beta R), \qquad
m_e = \tanh \left(\beta J/4 \right) m_{{\rm IM}} (\beta R), 
\label{mag}
\end{eqnarray}  
where $m_{\rm IM} (\beta R)$ denotes the spontaneous magnetisation of the spin-1/2 Ising model 
on the corresponding undecorated lattice that is known for several planar lattices \cite{lin92}. 

\section{Results and discussion}

Let us proceed to a discussion of the most interesting results obtained in the preceding section. First, it should be realized that the effective coupling (\ref{mp}) of the spin-1/2 Ising 
model on the corresponding undecorated lattice is always positive, i.e. $\beta R > 0$, which means 
that the interacting spin-electron system is effectively mapped to the ferromagnetic spin-1/2 
Ising model. Hence, it directly follows from Eq.~(\ref{mag}) that the ground state constitutes 
either the classical ferromagnetic phase with fully saturated and identically oriented 
sublattice magnetisations $m_i = m_e = 1/2$ on assumption that $J>0$, or the classical 
ferrimagnetic phase with fully saturated sublattice magnetisations oriented opposite 
one to each other $m_i = -m_e = 1/2$ if $J<0$. Both the spontaneously long-range ordered 
phases exhibit completely the same critical behaviour as a result of the invariance 
of the effective coupling $\beta R$ with respect to the transformation $J \to -J$.

\begin{figure}[h]
\begin{minipage}{8cm}
\includegraphics[width=8cm]{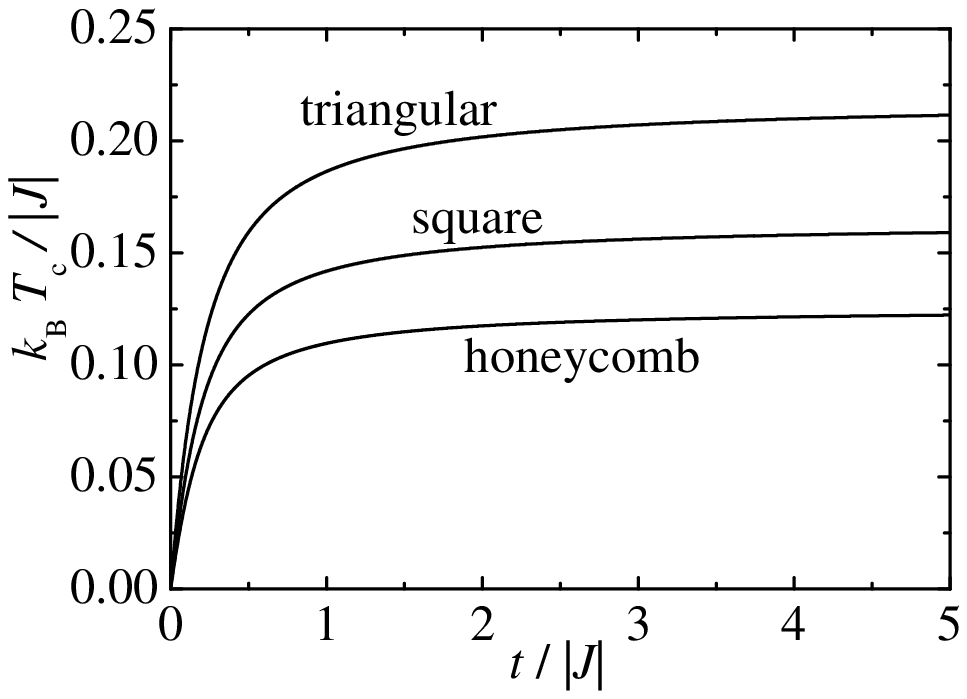}
\vspace{-1.1cm}
\caption{\label{fig2}The dimensionless critical temperature as a function of the relative strength 
of the kinetic term for doubly decorated honeycomb, square, and triangular lattices.}
\end{minipage}\hspace{0.5cm}%
\begin{minipage}{7.7cm}
\vspace{0.25cm}
\includegraphics[width=7.7cm]{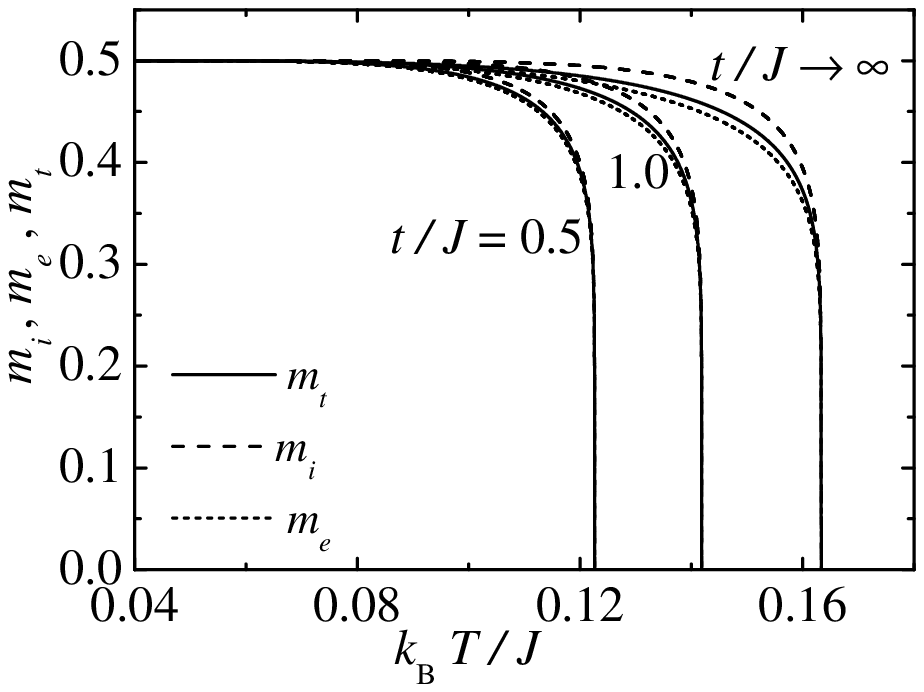}
\vspace{-1.2cm}
\caption{\label{fig3}Temperature variations of the total and both sublattice magnetisations 
for the doubly decorated square lattice at three different values of the ratio $t/J$.}
\end{minipage} 
\end{figure}

The reduced critical temperature of the spontaneously long-range ordered phases is plotted 
in Fig.~\ref{fig2} against the ratio $t/|J|$ between the hopping term and the exchange integral. 
As one can see, the critical temperature rises steadily when increasing a relative strength of 
the kinetic term until it asymptotically tends towards its maximum value achieved in the limit 
$t/|J| \to \infty$. In the limit $t/|J| \to 0$, the observed zero critical temperature  is 
consistent with a non-magnetic character of one of two decorating sites on each bond of 
the doubly decorated lattice. On the other hand, the highest asymptotic values of the critical 
temperatures for the interacting spin-electron system on the doubly decorated honeycomb, 
square and triangular lattices
\begin{eqnarray}
\frac{k_{\rm B} T_{\rm c}}{J_{\rm hc}} 
          = \frac{1}{4 \ln (2 + \sqrt{3} + \sqrt{6 + 4 \sqrt{3}} )}, 
\frac{k_{\rm B} T_{\rm c}}{J_{\rm sq}} 
          = \frac{1}{4 \ln (1 + \sqrt{2} + \sqrt{2 + 2 \sqrt{2}} )},
\frac{k_{\rm B} T_{\rm c}}{J_{\rm tr}}  = \frac{1}{4 \ln (\sqrt{2} + \sqrt{3})},
\label{tc}
\end{eqnarray}  
are exactly a half of the relevant critical temperatures of the spin-1/2 Ising model on singly decorated honeycomb, square, and triangular lattices, respectively. This observation would suggest 
that a delocalization of itinerant electrons generally reduces the critical temperature, because 
the localized Ising spins effectively feel due to the hopping process less than a half of 
the total magnetic moment of each itinerant electron. 

Thermal variations of the total and sublattice magnetisations are displayed in Fig.~\ref{fig3}
for three different values of a relative strength of the kinetic term. It is quite obvious from 
this figure that both sublattice magnetisations exhibit quite similar temperature dependencies
when the sublattice magnetization $m_i$ of the localized Ising spins lies just slightly above
the sublattice magnetisation $m_e$ of the itinerant electrons. A closer analysis reveals 
that both sublattice magnetisations tend to zero in a vicinity of the critical temperature 
with the critical exponent from the standard Ising universality class.

To conclude, we have found the exact solution for the hybrid model of the interacting spin-electron 
system on doubly decorated 2D lattices. Under the assumption of  quarter filling of each couple 
of the decorating sites, the ground state constitutes either spontaneously long-range ordered ferromagnetic or ferrimagnetic phase in dependence on whether the ferromagnetic or antiferromagnetic interaction between the localized Ising spins and itinerant electrons is assumed. 
It has been shown that the critical temperature of spontaneously long-range ordered phases monotonically increases upon strengthening the ratio between the kinetic term and the exchange interaction.  The work on an analogous 2D hybrid model of interacting spin-electron system with 
two electrons per each couple of the decorating sites is in progress \cite{stre09}.

\ack{This work was supported by the Slovak Research and Development Agency under the contract 
LPP-0107-06 and by Ministry of Education of SR under the grant VEGA~1/0128/08.}


\section*{References}
\begin{thebibliography}{9}

\bibitem{wu09}
Wu F Y 2009 \textit{Exactly Solved Models: A Journey in Statistical Mechanics} 
(Singapore: World Scientific)

\bibitem{fish59} 
Fisher M E 1959 \textit{Phys. Rev.} \textbf{113}  969

\bibitem{roja09} 
Rojas O, Valverde J S, de Souza S M 2009 Physica A \textbf{388} 1419

\bibitem{pere08} 
Pereira M S S, de Moura F A B F, Lyra M L 2008 \textit{Phys. Rev. B} \textbf{77} 024402

\bibitem{pere09} 
Pereira M S S, de Moura F A B F, Lyra M L 2009 \textit{Phys. Rev. B} \textbf{79} 054427

\bibitem{lin92} 
Lin K Y 1992 \textit{Chinese J. Phys.} \textbf{30} 287

\bibitem{stre09} 
Stre\v{c}ka J, Tanaka A, \v{C}anov\'a L, Verkholyak T, in preparation.

\end{thebibliography}
\end{document}